\RequirePackage{fix-cm}
\documentclass[twocolumn,epjc3]{svjour3}

\usepackage[numbers,sort&compress]{natbib}

\usepackage{graphicx}
\usepackage{xcolor}
\usepackage{amsmath}
\usepackage{url}      
\usepackage[colorinlistoftodos]{todonotes}
\usepackage{etex}
\usepackage{tikz}
\usepackage{tikz-feynman}
\PassOptionsToPackage{hyphens}{url}

\usepackage{xurl}   
\usepackage{hyperref}  
\usepackage{etoolbox}    
\apptocmd{\thebibliography}{\sloppy\setlength{\emergencystretch}{3em}}{}{}
\usetikzlibrary{positioning,calc}

\journalname{Eur. Phys. J. C}

\begin{document}

\title{Techniques for mass peak reconstruction in Searches for Long-Lived Heavy Neutral Leptons Decaying to a Lepton and a
\(\rho\) Meson
}

\titlerunning{Mass peak reconstruction for long-lived HNLs.}

\author{Marzieh Bahmani\thanksref{e1,addr1}
        \and
        Alessandro Guida\thanksref{e2,addr1}
        \and 
        Maral Khandan \thanksref{e3,addr2}
        \and 
        Heiko Markus Lacker\thanksref{e4,addr1}
        \and
        Anupama Reghunath \thanksref{e5,addr1}
}

\thankstext{e1}{e-mail: mbahmani@physik.hu-berlin.de}
\thankstext{e2}{e-mail: guida@physik.hu-berlin.de}
\thankstext{e3}{e-mail: marak92@zedat.fu-berlin.de}
\thankstext{e4}{e-mail: lacker@physik.hu-berlin.de}
\thankstext{e5}{e-mail: anupama.reghunath@physik.hu-berlin.de}

\institute{Institute of Physics, Experimental Physics department, Humboldt University of Berlin, Germany \label{addr1}
           \and
           Institute of Physics, Free University of Berlin, Germany \label{addr2}
}

\date{Received: date / Accepted: date}

\maketitle
\begin{abstract}
\begin{sloppypar}
The precise reconstruction of the mass peak of long-lived heavy neutral leptons (HNLs) helps to improve the sensitivity for sterile neutrino searches in both fixed-target and collider environments (e.g., SHiP and the LHC). We present an analytical framework for reconstructing the HNL mass peak in the semileptonic HNL decay channel \(N\rightarrow l\rho\) with \(\rho\to\pi\pi^0\), using only the lepton and the charged pion emerging from the decay vertex together with kinematic constraints and the known particle masses. Incorporating mass constraints from intermediate resonances (e.g., the \(\rho\) meson) or the parent particle (e.g., the \(W\) boson at the collider experiments), we propose two methods, suitable for experiments with displaced vertex tracking capabilities. 
The particle-level simulation's results demonstrate that the \(\rho\)-mass constraint method yields promising HNL mass resolution in both beam-dump and collider-based environments. The W-mass constraint method, limited to the HNLs produced via \(W\)-boson decays at the collider-based experiments, shows better HNL mass resolution than the \(\rho\)-mass constraint method.
\end{sloppypar}
\end{abstract}

\keywords{Heavy Neutral Leptons (HNLs) \and Semileptonic Decay \and SHiP \and LHC \and mass reconstruction}

\section{Introduction}
\label{intro}
\begin{sloppypar}
 Despite its successes and internal consistency, the Standard Model (SM) does not account for several key observations, such as the neutrino oscillations, which imply non-zero neutrino masses, the matter-antimatter asymmetry in the universe, and the existence of dark matter. 
 These limitations motivate the search for physics beyond the Standard Model (BSM), including theories that propose the existence of HNLs, which could provide solutions to some or even all of these fundamental problems~\cite{shaposhnikov2007}.
In the framework of the SM, neutrinos are considered to be massless, with strictly flavor-conserving interactions. However, the observation of neutrino flavor oscillations~\cite{esteban2020} demonstrates that neutrinos are massive.
Neutrino masses could be explained via the type-I seesaw mechanism, which introduces right-handed majorana neutrino states (HNLs) that are singlets under the SM gauge group. These HNLs mix with the SM neutrinos, enabling them to participate indirectly in weak interactions.
 Below the electroweak scale, the mixing between HNLs and active neutrinos is typically very small, which strongly suppresses their decay width~\cite{gronau1984}. This results in HNLs having macroscopic lifetimes, causing them to travel a measurable distance before decaying. Such long-lived HNLs can produce distinct displaced decay vertices in the detector, a striking experimental signature that helps suppressing many SM backgrounds.

These displaced decays provide a unique experimental insight to search for and characterize HNLs. By reconstructing the kinematics of the decay products, searches can directly constrain the HNL mass and mixing parameters, enabling tests of both the seesaw mechanism and models of sterile neutrinos across a broad range of parameter space.
Several experimental programs have placed significant constraints on the HNL parameter space across a wide mass range. 
In the multi-GeV regime, dedicated searches by ATLAS and CMS provide the most stringest limits on active–sterile mixing~\cite{ATLAS2023,CMS2022}.
In the sub-GeV mass range, experiments such as NA62 have recently searched for the decay $K\rightarrow \ell N$ and set upper limits on its branching fraction, thereby directly constraining HNL production in kaon decays~\cite{NA62HNL2021}. A complementary search technique at running and future experiments is offered by HNL mass-peak reconstruction in stopped charged-kaon decay events~\cite{Alves_2025}. 
Additionally, fixed-target experiments like PS191 and CHARM have been reinterpreted in modern global analyses~\cite{Drewes2018} to constrain active-sterile mixing and mass range up to 2 GeV. 
The future experiment at a dedicated beam-dump facility at CERN, SHiP, is expected to significantly extend the search sensitivity in the mass range 0.5-5 GeV to smaller mixing angles~\cite{SHIP2016}. 
In experimental searches, the reconstructed or inferred HNL mass is used as the final discriminant to enhance signal sensitivity and suppress background contamination. An accurate determination of the HNL mass plays a central role in setting exclusion limits and defining the sensitivity reach of each analysis across different masses and lifetimes. 

In the low-mass regime (below a few GeV), semileptonic decays
\(N\!\to\!\ell h\) with \(h\in\{\pi,\rho\}\) have sizable branching fractions and are
well suited to displaced-vertex (DV) searches. The pseudoscalar mode
\(N\!\to\!\ell\pi\) \footnote{Throughout all the equations, \(N\) denotes the HNL.} yields a clean two-track DV
topology in which the invariant mass \(m_{\ell\pi}\) peaks at \(m_{N}\) (HNL mass), providing strong
signal–background separation. The vector mode \(N\!\to\!\ell\rho\) with  \(\rho\!\to\!\pi\pi^{0}\) also features a visible
lepton and a charged pion, and its branching fraction is often comparable to or
larger than \(N\!\to\!\ell\pi\), in parts of the few-GeV region (model dependent). 
However the visible invariant mass \(m_{\ell\pi}\) deviates from the \(m_{N}\) peak due to the missing \(\pi^0)\).
This complication arises because  \(\pi^{0}\!\to\!\gamma\gamma\) nearly \(100\%\) of the time; without
calorimetric information the \(\pi^{0}\) is undetected, inducing missing momentum and
smearing the reconstructed mass. Therefore, fully exploiting \(N\!\to\!\ell\rho\) requires robust kinematic
reconstruction techniques to compensate for the undetected \(\pi^{0}\). It is important to note that in
hadron collider environments, residual backgrounds (e.g.\ heavy-flavour decays) are non-negligible for both \(N\!\to\!\ell\pi\)~\cite{ATLAS2025} and
\(N\!\to\!\ell\rho\) decay channels.
In this mass range, the purely leptonic three-body decays \(N\!\to\!\ell^\pm\ell^\mp\nu\) can become competitive or dominant
as phase space opens, but they lead to different experimental topologies. 
 
This study presents a compact analytic approach to HNL mass reconstruction: rotate the coordinates to align with the HNL flight direction. By combining this rotated frame approach with kinematic constraints from the known masses of intermediate particles (e.g. the \(\rho\) meson or the parent W boson), the \(\pi^{0}\)\ longitudinal momentum can be estimated. 
In this paper, we present and compare two kinematic reconstruction methods aimed at estimating the mass of long-lived HNLs decaying into the semileptonic final state \(N\rightarrow l\rho\) without relying on the \(\pi^0\) reconstruction. Section~\ref{sec:Theory} provides the theoretical and phenomenological context for HNL production and decay, highlighting the relevance of semileptonic modes, and introducing the experimental challenges in reconstructing the \(N\rightarrow l\rho\) decay channel.
Section~\ref{sec:Kinematic} introduces a rotated coordinate system aligned with the HNL flight direction, which simplifies the estimate of the missing momentum. Section~\ref{sec:rho-const} details the \(\rho\)-mass constraint method. Section~\ref{sec:case-rho} presents performance evaluations using simulations from both beam-dump and collider-based environments. Section~\ref{W-conts-method} explores a complementary method that utilizes the W-boson mass as a constraint, applicable to collider-based HNL production. As an outlook, we propose an adaptation of our resonance–constraint strategy to HNLs coupling to \(\tau\) leptons in Section~\ref{tau-coupling}.
 We conclude our findings in Sections~\ref{conclusion}  and discuss the effectiveness and applicability of each reconstruction method under different experimental conditions.
\end{sloppypar}
\section{Theoretical and Phenomenological Context}
\label{sec:Theory}
\begin{sloppypar}
Understanding how HNLs are produced and decay is essential for designing effective strategies in experiments searching for long-lived particles. This section provides an overview of the dominant production mechanisms, typical decay channels, and the experimental signatures of HNLs in both fixed-target and collider settings.
Due to their suppressed mixing with active neutrinos, HNLs often exhibit long lifetimes, leading to decays that occur at macroscopic distances from the production point. This gives rise to a characteristic DV signature, which is a key feature in experimental searches for HNLs. 
Beam-dump and collider experiments probe complementary regions of the HNL mass–lifetime parameter space. Beam-dump facilities are most sensitive to long lifetimes, but their accessible mass range is limited by the available production mechanisms in heavy-flavour meson decays, since the center-of-mass energy is too low for on-shell W production. Collider experiments, by contrast, can access higher HNL masses through W/Z boson and heavy-flavour decays, and since the decay volume is smaller are thus more sensitive to shorter lifetimes.
\end{sloppypar}
\subsection{HNL Production in Beam-Dump and Fixed-Target Experiments}
\begin{sloppypar}
In non-collider environments, such as beam dump and fixed-target experiments (e.g., SHiP, DUNE, NA62), HNLs are primarily produced via the decays of light mesons (\(\pi\),\(K)\) as well as heavy flavour mesons~\cite{SHIP2019}, as a result of the high-intensity proton beams striking a dense target. We focus here on heavy mesons including leptonic and semileptonic meson decays such as:
\begin{align*}
    D^\pm, D^{\pm}_{s}, B^\pm \rightarrow \ell^\pm N,  \quad B \rightarrow X_c \ell N ,D \rightarrow K\,\ell N\,.
\end{align*}
 these processes leverage the large meson yields produced in such experiments and are  sensitive to HNLs with masses between 0.5 and 5 GeV.
These setups are ideal for exploring long-lived particles due to their large decay volumes and minimal backgrounds. The long lifetimes of HNLs, stemming from their small mixing with active neutrinos, allow decays to occur well downstream of the target, making them prime candidates for searches with DV signatures.
\end{sloppypar}
\subsection{HNL Production at Colliders}
At high-energy colliders, such as the LHC or future lepton colliders, HNLs can be produced through the decay of electroweak gauge bosons. One of the main production modes at proton-proton colliders is:
\begin{align*}
    W^\pm \rightarrow \ell^\pm N
\end{align*}
This process yields a prompt charged lepton, which provides a useful experimental signature to trigger on and reconstruct the event.
In contrast, at lepton colliders, HNLs can be produced through:
\begin{align*}
    e^+ e^- \rightarrow Z \rightarrow \nu N
\end{align*}
Although this channel lacks a prompt lepton, it also avoids the hadrons/jets coming from the interaction point, typical of hadron colliders, reducing background and offering a cleaner environment for HNL searches.

\subsection{HNL Decay Channels}
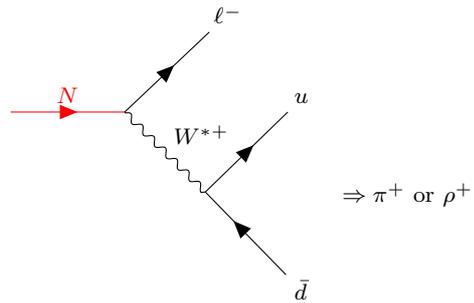
\begin{figure}[htbp]
\centering
\begin{tikzpicture}
  \begin{feynman}
    \vertex (v1); 
    \vertex [right=of v1] (hnl);
    \vertex [above right=of hnl] (lep) {\(\ell^-\)};
    \vertex [below right=of hnl] (v2);
    \vertex [above right=of v2] (u) {\(u\)};
    \vertex [below right=of v2] (dbar) {\(\bar{d}\)};
    
    \diagram* {
      (v1) -- [fermion, red, edge label=\(N\)] (hnl),
      (hnl) -- [fermion] (lep),
      (hnl) -- [boson, edge label=\(W^{*+}\)] (v2),
      (v2) -- [fermion] (u),
      (v2) -- [anti fermion] (dbar),
    };
  \end{feynman}
  \node at ($(u)!0.5!(dbar) + (1.4,0)$) {\(\Rightarrow \pi^+ \text{ or } \rho^+\)};
\end{tikzpicture}
\caption{Quark-level decay of an HNL: \(N \rightarrow \ell^- + W^{*+} \rightarrow \ell^- + u\bar{d}\), where the \(u\bar{d}\) hadronizes into either a \(\pi^+\) (pseudoscalar) or a \(\rho^+\) (vector meson), which eventually decays into a $\pi^+$ and \(\pi^0\)}
\label{fig:N_to_lep_rho_or_pion}
\end{figure}

\begin{sloppypar}
In the few-GeV mass range, HNLs can decay through both leptonic three-body and semileptonic two-body channels. The purely leptonic modes
\(N \to \ell^{+}\ell'^{-}\nu\) become dominant around \(m_N\simeq 1.5\text{–}2~\mathrm{GeV}\) as phase space opens up~\cite{Bondarenko2018}; they offer clean experimental signatures (especially for \(e/\mu\)) with branching fractions that depend on \(m_N\) and the active–sterile mixing pattern. Semileptonic modes \(N \to \ell\,h\) with a light meson \(h\in\{\pi,\rho,\ldots\}\) are also relevant. In this work, we focus on the vector-meson channel
\(N \to \ell\,\rho\) with \(\rho^{+}\to\pi^{+}\pi^{0}\) and \(\ell\) being \(e^\pm\) or \(\mu^\pm\) \footnote{Unless stated otherwise, charge-conjugated decays \(N\to \ell^{+}\rho^{-}\) are included.}. At quark level, this proceeds via \(N \to \ell^{-}W^{*+}\to \ell^{-}u\bar d\), where hadronization yields either a pseudoscalar \(\pi^{+}\) or a vector \(\rho^{+}\) (see Fig.~\ref{fig:N_to_lep_rho_or_pion}).
\end{sloppypar}

 Experimentally, the \(\rho\) channel produces a DV with a lepton and a charged pion.
The charged pion leaves a visible track in the detector, but the neutral pion decays almost immediately into two photons with almost \(100\%\) branching ratio, $\pi^0 \rightarrow \gamma\gamma$.
In most detectors with a tracking system, charged pions as well as light leptons are directly reconstructed. However, the \(\pi^{0}\) momentum must be inferred from calorimeter information (where applicable).
Moreover, the photons carry only a fraction of the HNL energy and hence might have too low energy to be efficiently detected.
If the \(\pi^0\) escapes direct detection, this results in a partially reconstructed final state. 
In the following sections, we introduce a novel mass reconstruction techniques that  enables a reliable reconstruction of the HNL mass even without detecting the \(\pi^0\).
\section{Kinematic Framework}
\label{sec:Kinematic}
\begin{sloppypar}
The decay topology of interest features a DV in which an HNL decays into a charged lepton and a hadronic system. The direction of the HNL's flight, denoted by \(\hat{n}\), taken as the unit vector from the primary to the displaced vertex. In a DV-aligned frame, momentum conservation implies that the components of the $\pi^0$ momentum transverse to $\hat{n}$ satisfy
$p_{\pi^0}^{\,\perp} = -\sum_{\text{vis}} p_{i}^{\,\perp}$,
so the two transverse components are fixed and only the longitudinal component $p_{\pi^0}^{\parallel}$ remains unknown. We adopt this DV-aligned frame following the ATLAS strategy~\cite{ATLAS2023}. This frame streamlines the application of momentum conservation and provides a natural basis to impose additional kinematic constraints, such as the resonance-mass conditions discussed next.
\end{sloppypar}

\subsection{Rotated Frame Setup}
\begin{sloppypar}
To simplify the kinematic reconstruction of the HNL,  a new rotated coordinate system, denoted as $(\hat{\mathbf{x}}', \hat{\mathbf{y}}', \hat{\mathbf{z}}')$ is defined,
which is aligned with the HNL flight direction. 
The origin of this rotated frame is placed at the HNL production point, and the \(\hat{\mathbf{z}}'\)-axis points along the HNL momentum vector, \(\hat{n}\). The transverse and longitudinal components of the decay products' momenta are naturally separated in this frame, simplifying the application of kinematic constraints.
The rotated axes are:
\begin{equation}
\hat{\mathbf{z}}' = \frac{\hat{n}}{|\hat{n}|},\quad
\hat{\mathbf{x}}' =\hat{\mathbf{z}}' \times \hat{\mathbf{y}}',\quad
\hat{\mathbf{y}}' = \hat{\mathbf{z}}' \times \hat{\mathbf{x}}'
\end{equation}

\end{sloppypar}
\begin{sloppypar}
The determination of the HNL flight direction \(\hat{n}\) depends on the 
experimental environment. At the LHC, \(\hat{n}\) is reconstructed as the line 
connecting the primary vertex (PV), defined by the proton--proton interaction 
point, to the secondary vertex (SV) where the displaced HNL decay occurs,
\begin{equation}
\hat{n}_{\text{LHC}} = 
\frac{\vec{x}_{\text{SV}} - \vec{x}_{\text{PV}}}
     {|\vec{x}_{\text{SV}} - \vec{x}_{\text{PV}}|}\,.
\end{equation}
In contrast, at a beam-dump facility such as SHiP there is no well-defined PV 
within the detector, since HNLs are produced inside the target and absorber 
complex. Instead, the production point is taken to be the known position of the 
target, and $\hat{n}$ is defined as the vector from this location to the 
reconstructed decay vertex within the fiducial volume,
\begin{equation}
\hat{n}_{\text{SHiP}} = 
\frac{\vec{x}_{\text{SV}} - \vec{x}_{\text{target}}}
     {|\vec{x}_{\text{SV}} - \vec{x}_{\text{target}}|}\,.
\end{equation}
This procedure exploits the tight collimation of the proton beam, which provides a precise reference direction for the HNL production axis. In addition, the long baseline $|\vec{x}_{\text{SV}}-\vec{x}_{\text{target}}|$ is typically tens of meters at SHiP and therefore reduces the sensitivity of $\hat n$ to positional uncertainties.

\end{sloppypar}

\section{\(\rho\)-Mass Constraint Method}
\label{sec:rho-const}
In this method, we assume that the invariant mass of the charged pion and the missing neutral pion system corresponds to the nominal \(\rho\) meson mass. This assumption provides a kinematic constraint that helps reconstruct the invisible momentum vector of the \(\pi^0\). Importantly, this technique relies solely on the final-state kinematics of the HNL decay products and is therefore independent of its production mechanism. Hence, whether the HNL is produced in a collider, fixed-target, or beam-dump experiment, the applicability of this method remains unchanged.

\subsection{Quadratic Solution for Neutral Pion Longitudinal Momentum}
\begin{sloppypar}
As discussed in Section~\ref{sec:Kinematic}, after rotation, the HNL has no transverse momentum in this frame ($p^{\perp}_{N}=0$), therefore the vector sum of the daughters’ transverse momenta vanishes. This fully constrains the transverse momentum of \(\pi^{0}\)\ from the two measured tracks and leaves only its longitudinal component unknown.
In this rotated frame, the momentum of \(\pi^0\) is denoted by $\vec{q}$, and we can write the transverse momenta of \(\pi^0\) as
\begin{equation}
    q^{\perp} \equiv p^{\perp}_{\pi^0} = - (p^{\perp}_{\pi}+p^{\perp}_{l})
    \label{eq:tranM}
\end{equation}
and the longitudinal momenta as: \(q^{\parallel} \equiv p_{\pi^0}^{\parallel}\).
The four-momentum of the \(\rho\) meson is given by:
\begin{equation}
    P_{\rho}=P_{\pi}+P_{\pi^{0}}\,.
    \label{eq:rho4}
\end{equation}
Equation~\ref{eq:rho4} is squared and the Minkowski inner product identity  \(P^2 = E^2 - |\vec{p}|^2 = m^2\) is applied \footnote{c (speed of light) is set to 1.} to obtain:
\begin{align}
    m_{\rho}^{2} = m_{\pi}^2+m_{\pi^0}^2+2 E_{\pi} \sqrt{m_{\pi^0}^2+q^{\perp 2}+q^{\parallel 2}}\nonumber\\-2p^{\parallel}_{\pi}q^{\parallel}
    - 2q^{\perp} \cdot p^{\perp}_{\pi}\,.
    \label{eq:mrho}
\end{align}
Here, \(m_{\pi^0}\) and \(m_{\pi}\) are set to their known masses of the neutral and charged pion respectively. Their values are taken from Particle Data Group (PDG) reviews~\cite{pdg2022} \(m_{\pi^{0}}=134.9\) \(\mathrm{MeV}\), \(m_{\pi^{+}}=139.57\) \(\mathrm{MeV}\). The mass of the \(\rho\) meson is also taken from PDG \(m_{\rho} = 775.26\) \(\mathrm{MeV}\). 
Transforming Equation~\ref{eq:mrho} into a quadratic equation in \(q^{\parallel}\)\ results in:
\begin{equation}
a\,q^{\parallel 2}
+ b\,q^{\parallel}
+ c = 0
\tag{8}
\label{eq:mrho_quad}
\end{equation}
where the coefficients are given by:
\begin{equation}
a = E_{\pi}^2 - p_{\pi}^{\parallel 2}
\tag{8a}
\label{eq:mrho_a}
\end{equation}

\begin{equation}
b = p_{\pi}^{\parallel}\Bigl(
        m_{\pi}^2 + m_{\pi^0}^2 - m_{\rho}^{2}
        - 2\,p_{\pi}^{\perp}\cdot q^{\perp}
    \Bigr)
\tag{8b}
\label{eq:mrho_b}
\end{equation}

\begin{equation}
c = E_{\pi}^2\Bigl(
        m_{\pi^{0}}^2 + q^{\perp 2}
    \Bigr)
    - E_{\pi}^2\Bigl(
        m_{\rho}^{2}
        - m_{\pi}^2
        - m_{\pi^0}^2
        - 2\,\bigl(p_{\pi}^{\perp}\cdot q^{\perp}\bigr)^2
    \Bigr)\,.
\tag{8c}
\label{eq:mrho_c}
\end{equation}
The solutions for Equation~\ref{eq:mrho_quad} are:
\begin{equation}
q^{\parallel}_{\text{max/min}} =
\frac{
   -b \,\pm\, \sqrt{\,b^2 - 4\,a\,c\,}
}{
   2\,a
}
\tag{9}
\label{eq:mrho_solution}
\end{equation}
The $+$ and $-$ signs in front of the square root corresponds to 
\(q^{\parallel}_{\text{max}}\) and \(q^{\parallel}_{\text{min}}\) respectively.

The four-momentum of the HNL in the decay \(N \rightarrow \ell \rho\) is given by:
\begin{equation}
    P_{N} = P_{\ell} + P_{\rho} 
    \tag{10}
    \label{eq-14}
\end{equation}
By squaring both sides of Equation~\ref{eq-14} and using Equation~\ref{eq:rho4} one obtains:
\begin{equation}
\begin{aligned}
m_{N}^{2} &= (P_{\ell} + P_{\rho})^2 =m_{\ell}^2 + m_{\rho}^2+ 2 P_{\ell} \cdot (P_{\pi} + P_{\pi^0})\,.
\end{aligned}
\tag{11}
\end{equation}
Expanding the dot product explicitly and applying  the Minkowski inner product identity, results in:
\begin{equation}
m_{N}^{2} =
m_{\ell}^2 + m_{\rho}^2 + 2E_{\ell} E_{\pi} - 2\vec{p}_{\ell} \cdot \vec{p}_{\pi} + 2E_{\ell} E_{\pi^0} \\
- 2\vec{p}_{\ell} \cdot \vec{p}_{\pi^0}
\tag{12}
\end{equation}
where \(E_{\ell}\), \(m_{\ell}\) and \(\vec{p}_{\ell}\) are the total energy, rest mass and momentum of the lepton, respectively.

The HNL mass can then be expressed as:
\begin{equation}
\begin{aligned}
m_{N}^{2} &= m_{\ell}^2 + m_{\rho}^2 + 2E_{\ell} E_{\pi} + 2E_{\ell} \sqrt{m_{\pi^0}^2 + q^{\perp 2} + q^{\parallel 2}} \\
          &\quad - 2 p_{\ell}^{\parallel}(p_{\pi}^{\parallel} + q^{\parallel} ) - 2 p_{\ell}^{\perp} \cdot p_{\pi}^{\perp}
\end{aligned}
\tag{13}
\label{eq:m_N}
\end{equation}

Substituting the two solutions \(q^{\parallel}_{\text{min/max}}\) from
Eq.~(\ref{eq:mrho_solution}) into Eq.~(\ref{eq:m_N}) yields two kinematic
branches for \(m_N^2\). Defining \(m_N^{\rho}\equiv\sqrt{m_N^2}\), we retain
only the positive root as the physical solution. We refer to the two results as
\(m_{N}^{\rho,\text{min}}\) and \(m_{N}^{\rho,\text{max}}\), corresponding to
the \(q^{\parallel}_{\text{min}}\) and \(q^{\parallel}_{\text{max}}\) inputs, respectively.

\end{sloppypar}
\begin{sloppypar}
When solving for the missing longitudinal momentum of the \(\pi^0\) (\(q^{\parallel}\)) using the \(\rho\)-mass constraint method, there are cases in which no real solution exists. This occurs when the argument of the square root as defined in Equation~\ref{eq:mrho_solution} becomes negative, indicating a kinematically incompatible configuration.
This incompatibility can be understood in light of the \(\rho\) meson's resonance nature. As an intermediate state in the decay \(N \to \ell \rho\), the \(\rho\) is not a stable particle but a resonance characterized by a Breit-Wigner mass distribution with an intrinsic width \(\Gamma_\rho\) of 149 MeV~\cite{pdg2022}. While the nominal \(\rho\) mass in this paper is used as a fixed value, the effective mass can fluctuate on an event-by-event basis within this natural width.
To determine the kinematically allowed range of the \(\rho\) meson invariant mass, we study \(m_\rho^2\) as a function of the unknown longitudinal momentum \(q^{\parallel}\) of the neutral pion. Since \(m_\rho^2(q^{\parallel})\) is quadratic in \(q^{\parallel}\), it possesses a minimum value. The location of this minimum is obtained by differentiating Equation~\ref{eq:mrho} with respect to \(q^{\parallel}\) and setting the derivative to zero,
\begin{equation}
    \frac{d m_\rho^2}{d q^{\parallel}} = 0.
    \tag{14}
\end{equation}
The solution yields the neutral pion longitudinal momentum at which the $\pi^+\pi^0$ invariant mass reaches its minimum,
\begin{equation}
    q^{\parallel}_{\text{min}} = \frac{p^{\parallel}_{\pi} \, E_{\pi^0}}{E_{\pi}}.
    \label{eq:qpar_solution}
    \tag{15}
\end{equation}
This minimum defines the lowest possible value of \(m_\rho\) compatible with the measured lepton and charged pion kinematics. 
The corresponding minimum \(\rho\)-mass for the event, \(m_{\rho,\text{min}}\), is then
\begin{equation}
    m_{\rho,\text{min}}^2 = m_{\pi}^{2} + m_{\pi^{0}}^{2} + 2E_{\pi} E_{\pi^{0}} - 2p_{\pi}^{\parallel} q^{\parallel}_{\text{min}} - 2 p_{\pi}^{\perp} \cdot q^{\perp} 
    \tag{16}
\end{equation}
For events in which the radicand in Eq.~\ref{eq:mrho_solution} is negative when using the nominal \(m_\rho\), we replace \(m_\rho \) by \(m_{\rho,\text{min}}\) and proceed with the reconstruction. This choice guarantees a real solution by construction.
\end{sloppypar}
\section{Performance Evaluation of the \(\rho\)-Mass Constraint Method in Beam-Dump and Collider Scenarios}
\label{sec:case-rho}
\begin{sloppypar}
To assess  the \(\rho\)-mass constraint method introduced in Section~\ref{sec:rho-const}, we evaluate its performance across two representative experimental environments: (i) a fixed-target beam-dump scenario, exemplified by the SHiP experiment, and (ii) a high-energy proton-proton collider scenario, such as the LHC. These two classes of experiments differ significantly in terms of HNL production modes, event topologies, and detector geometries.
The objective is twofold: first, to verify whether the kinematic constraint method is able to reconstruct a mass peak as closely as possible to the truth HNL mass and also of smaller width than the simple invariant mass \(m_{\ell\pi}\); and second, to test its discriminatory power in differentiating these from \(N \to \ell\pi \)  that yields the same charged particles final state but lacks the additional \(\pi^0\). This section presents simulation-based studies for each experimental scenario at particle level, hence without the effects from detector response.
  \end{sloppypar}
\subsection{Fixed-Target Scenario}
\begin{figure}[htbp]
    \centering
    \includegraphics[width=0.53\textwidth]{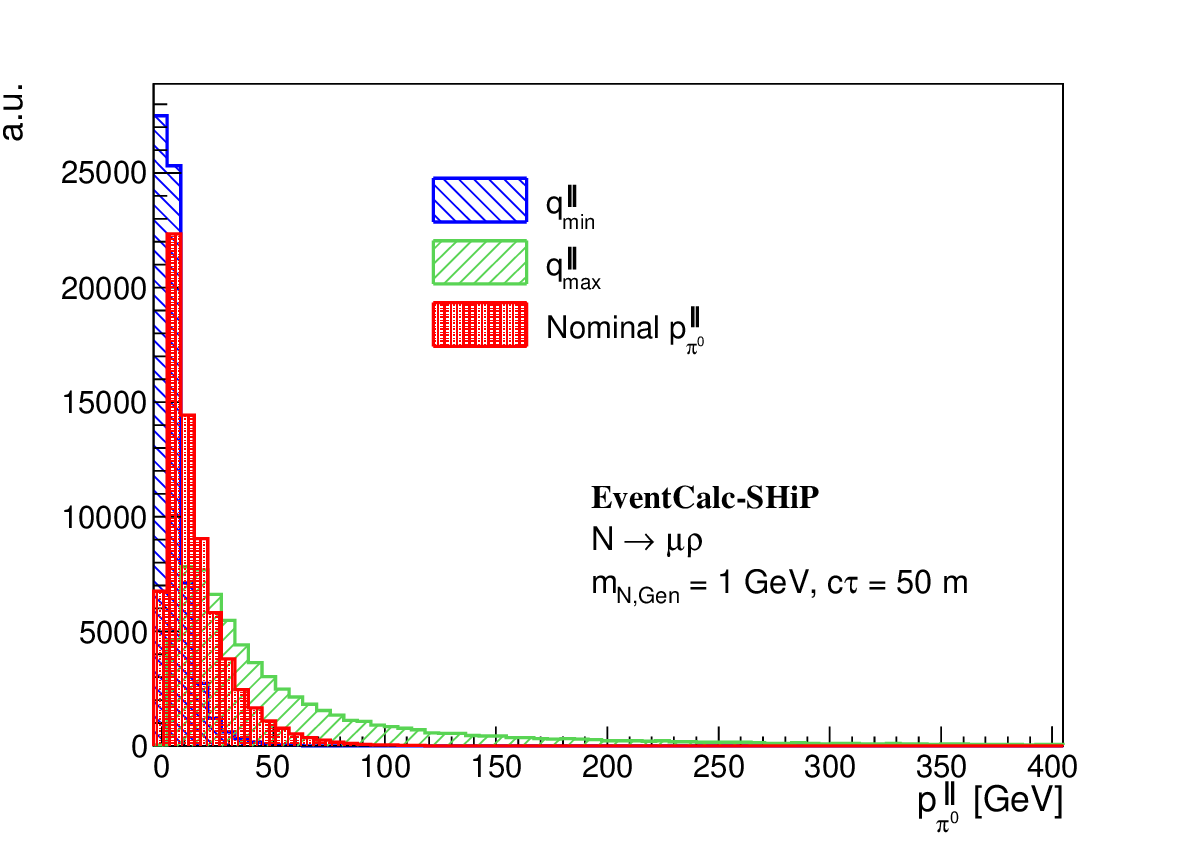}
    \caption{Longitudinal neutral–pion momentum \(p_{\pi^{0}}^{\parallel}\).
The filled (red) histogram shows the nominal (particle level), while the hatched (blue/green)
histograms show the two kinematic solutions \(q_{\min}^{\parallel}\) and
\(q_{\max}^{\parallel}\) obtained with the \(\rho\)-mass constraint method.
The sample corresponds to \(N\!\to\!\mu\rho\) with
\(m_{N,GeN}=1~\mathrm{GeV}\) and \(c\tau=50~\mathrm{m}\), generated with
\texttt{EventCalc--SHiP} in a SHiP-like configuration. The
\(q_{\min}^{\parallel}\) solution is softer and peaks at low momentum, whereas
\(q_{\max}^{\parallel}\) is broader and extends to higher momenta.}
    \label{fig:pion_pp_rho_method_SHiP}
\end{figure}

\begin{sloppypar}
To evaluate the proposed technique in a fixed-target context, signal samples specifically tailored to the SHiP experiment are employed. These samples are generated using \texttt{EventCalc-SHiP}~\cite{EventCalcSHiP}, a custom event generator developed for long-lived particle studies at SHiP. Within this framework, the decay products are subsequently processed using \texttt{Pythia8}~\cite{pythia8}. 
\end{sloppypar}
\begin{sloppypar}
Two heavy neutral lepton (HNL) decay channels are considered: 
\(N \rightarrow \mu \rho\) and \(N \rightarrow \mu \pi\), both simulated for a 
benchmark HNL mass of $1\,\mathrm{GeV}$ and a proper decay length of \(50\,\mathrm{m}\). 
The choice of \(1\,\mathrm{GeV}\) ensures that the decay 
\(N \to \mu \rho\) is kinematically allowed, i.e.\ \(m_N > m_\rho + m_\mu\), while 
remaining below \(2\,\mathrm{GeV}\), where the SHiP experiment has its highest 
sensitivity from charm-meson production of HNLs. 
Moreover, the branching ratio for \(N \to \ell \rho\) decreases with increasing 
\(m_{N}\), making the \(1\,\mathrm{GeV}\) benchmark representative of a favorable 
region of parameter space for SHiP. While the \(\mu \rho\) channel features a 
genuine \(\rho^\pm\) resonance, both decay modes can yield identical reconstructed 
final states (\(\mu \pi\)) in events where the neutral pion from 
\(\rho^\pm \to \pi^\pm \pi^0\) is not detected. 
In the present study, the HNL is generated at a single point, corresponding to 
the nominal production vertex. In a realistic beam-dump environment, the 
effective production region is extended by the thickness of the target, which 
introduces a smearing of the true production point.
\end{sloppypar}
\begin{sloppypar}
Figure~\ref{fig:pion_pp_rho_method_SHiP} displays the two kinematic solutions for the longitudinal momentum of the reconstructed \(\pi^{0}\) in the \(\mu\rho\) sample, overlaid with the corresponding event-by-event simulated nominal values for comparison.
Note that the reconstructed-solution histograms do not, event-by-event match the nominal value. In shape, the minimum solution is slightly narrower and more peak-like, while the maximum solution is broader with longer tails. In an experimental setting there is no a priori best choice of the solutions. 

\begin{figure}[htbp]
    \centering
    \includegraphics[width=0.53\textwidth]{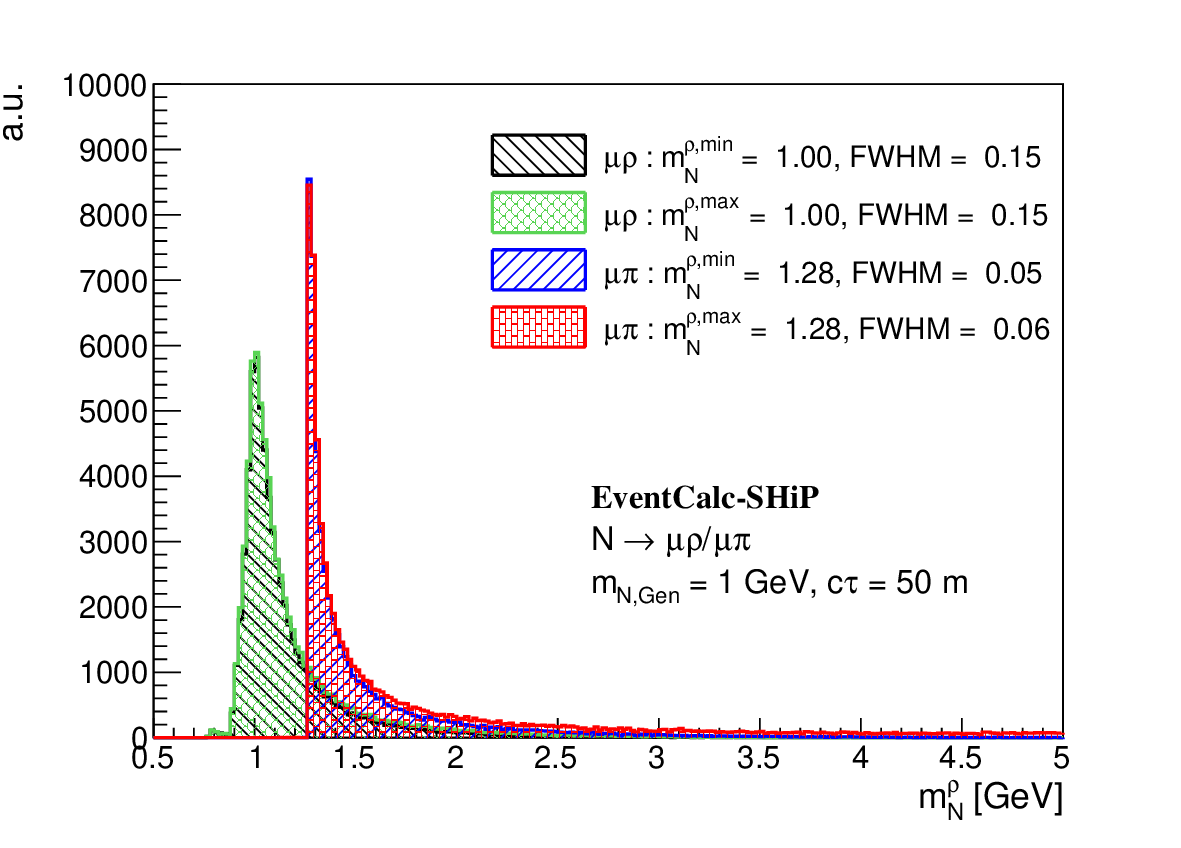}
    \caption{
    Reconstructed HNL mass \(m_{N}^{\rho,\text{min/max}}\) using the
\(\rho\)-mass constraint method for \(N\!\to\!\mu\rho\) and \(N\!\to\!\mu\pi\) samples
(\(m_{N,Gen}=1~\mathrm{GeV}\), \(c\tau=50~\mathrm{m}\)), generated with
\texttt{EventCalc--SHiP} in a SHiP-like configuration.
For \(N\!\to\!\mu\rho\), the two branches yield nearly identical spectra that peak at the
true mass, while \(N\!\to\!\mu\pi\) (with no \(\rho\) present) produces a displaced, narrower
peak at higher mass, allowing a clean separation of decay hypotheses.}

    \label{fig:hnl_mass_rho_method_SHiP}
\end{figure}

Figure~\ref{fig:hnl_mass_rho_method_SHiP} shows the \(m_{N}\) distributions reconstructed with the \(\rho\)-mass constraint method called \(m_{N}^{\rho, \text{min/max}}\), obtained from the two kinematic solutions, for both the \(\mu\rho\) and \(\mu\pi\) samples. The reconstructed HNL mass distributions \(m_{N}\) are statistically indistinguishable at particle level, so the peaks' position and resolution are insensitive to the choice of  \(q^{\parallel}_{\text{min/max}}\) solutions. Applying the \(\rho\)-mass constrained method uniformly across the two decay channels for the same benchmark \(m_{N,Gen}=1\)\,GeV results 
in two well-separated mass peaks for the respective samples, effectively 
distinguishing events that include a true \(\rho\) resonance from those that do 
not. For clarity, the relative branching fractions of the two channels are not 
taken into account in this figure, so only the kinematic separation is shown.
A complete overview of the Most probable  
values (MPV) and full-width at half maximum (FWHM) obtained is 
summarized in Table~\ref{tab:rho_constraint_summary}.
For the $\mu\rho$ sample, the reconstructed HNL mass peaks at 
$1.00\,\mathrm{GeV}$ with a FWHM of 
$0.15\,\mathrm{GeV}$, while for the $\mu\pi$ sample the peak is at 
$1.28\,\mathrm{GeV}$ with a narrower FWHM of $0.05\,\mathrm{GeV}$ for \(m_{N}^{\rho, \text{min}}\) and $0.06\,\mathrm{GeV}$  for \(m_{N}^{\rho, \text{max}}\) . 

Given the expected mass resolution of the SHiP experiment~\cite{SHiP:2022xqw,DeLellis:2018epjconf}, these 
results demonstrate that the reconstruction method provides both 
good accuracy for signal events and strong discrimination between \(N \to \ell \rho\) and \(N \to \ell \pi\). 
There are two additional handles to
distinguish between \(N \to \ell \rho\) and \(N \to \ell \pi\):

(i) The invariant mass distribution \(m_{\ell\pi}\) for \(N \to \ell \pi\) decay peaks at the true \(m_{N}\) value, while for the \(N \to \ell \rho\) decays, it is broadly distributed with its maximum value at the allowed kinematic endpoint, \(m_{N}\), due to the missing \(\pi^0\) momentum.

(ii) The impact parameter (IP) at the target~\cite{SHiP:2022xqw}, defined as the transverse distance between the target location and the reconstructed momentum vector of the visible final 
state, \((\vec{p}_\ell + \vec{p}_\pi)\). For genuine \(N \to \ell \pi\) decays, the IP 
values cluster near zero, while for \(N \to \ell \rho\) decays they are shifted to 
larger values, reflecting the fact that the \(\pi^0\) momentum is not included in 
the reconstruction. This observable therefore provides an additional 
discriminating feature that could enhance separation of the two channels in a 
realistic experimental analysis.

\end{sloppypar}


\subsection{Collider Scenario}
\label{sec:collidersim}
\begin{figure}[htbp]
    \centering
    \includegraphics[width=0.53\textwidth]{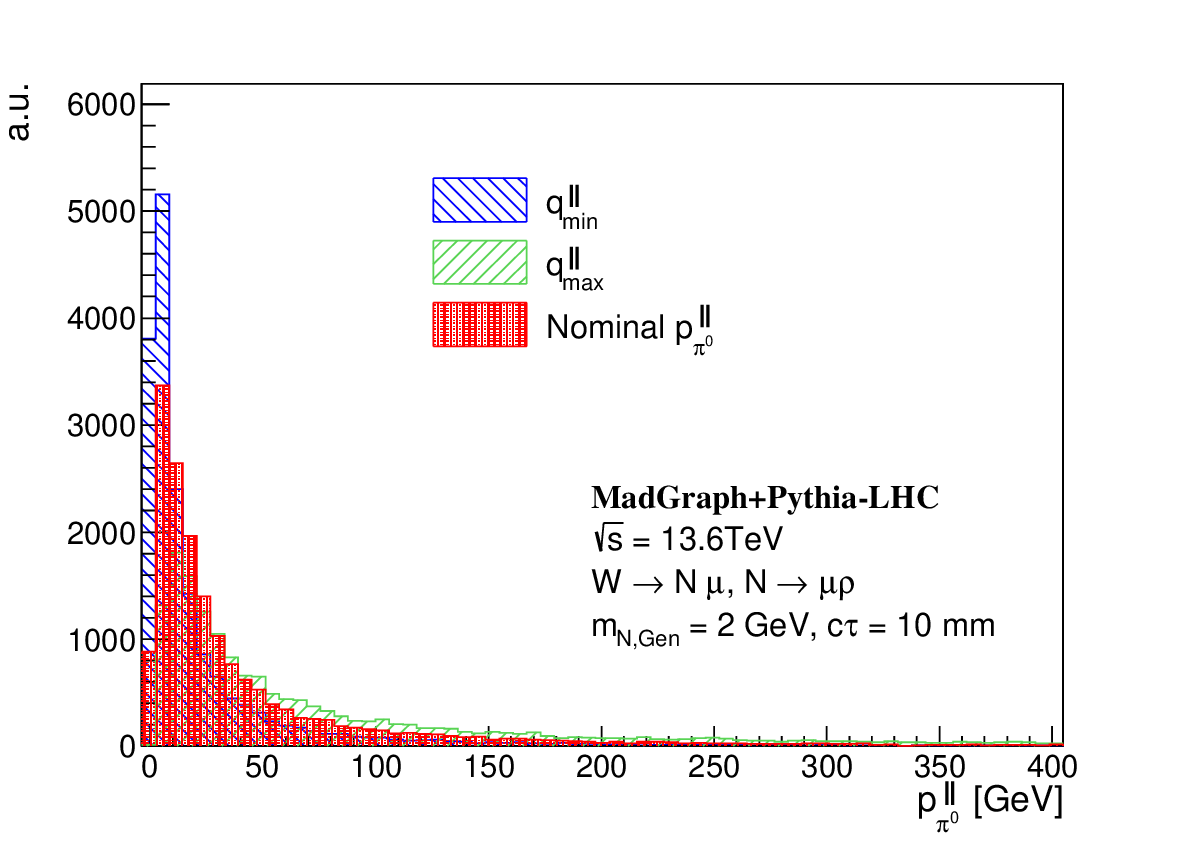}
    \caption{Longitudinal neutral pion momentum \(p_{\pi^{0}}^{\parallel}\) for an LHC-like sample with \(m_{N, Gen}=2~\mathrm{GeV}\), produced via
\(pp\!\to\! W^{\pm}\!+\!X \to \ell N\!+\!X\) and \(N\!\to\!\ell\rho\).
The filled histogram shows the nominal (particle level), while the hatched histograms correspond
to the two kinematic solutions \(q_{min/max}^{\parallel}\) obtained with the
\(\rho\)-mass constraint method. The
\(q_{\min}^{\parallel}\) solution is softer and peaks at low momentum, whereas
\(q_{\max}^{\parallel}\) is broader and extends to higher momenta.}
    \label{fig:pion_pp_rho_method}
\end{figure}

\begin{sloppypar}    
For the collider-based scenario, we simulate the production and decay of HNLs through the process \(pp \rightarrow W^{\pm} + X \rightarrow \ell N\) + X, \(N \rightarrow \ell \rho\). The signal events are generated using the \texttt{SM\_HeavyN\_meson\_NLO} model~\cite{Ruiz} in \texttt{MadGraph5\_aMC@NLO}~\cite{Madgraph}, which handles the hard scattering matrix element calculations. The parton shower, hadronization, and resonance decays, including those of the \(\rho\) meson, are performed using \texttt{Pythia8}~\cite{pythia8}, with the NNPDF30NLO~\cite{NNPDF} parton distribution functions applied.
The samples were generated assuming an HNL mass of \(2\,\mathrm{GeV}\) and a decay 
length of \(10\,\mathrm{mm}\) for the \(W^\pm \to \mu N\), \(N \to\mu\rho\). The choice of 
\(m_\text{N,Gen} = 2\,\mathrm{GeV}\) is motivated by the fact that for lower masses, existing 
limits from past beam-dump experiments already provide much stronger constraints, 
such that long-lived particle searches at ATLAS and CMS cannot realistically 
compete with. For this reason, we restrict our collider study to HNL masses above 
\(2\,\mathrm{GeV}\).
\end{sloppypar}
\begin{sloppypar} 
Figure~\ref{fig:pion_pp_rho_method} shows the longitudinal \(\pi^{0}\) momentum for the minimum and maximum solution according to the Equation~\ref{eq:mrho_solution} in comparison with the nominal simulated value from the generated sample. Similar behaviours as the SHiP-like scenario is observed.

\begin{figure}[htbp]
    \centering
    \includegraphics[width=0.53\textwidth]{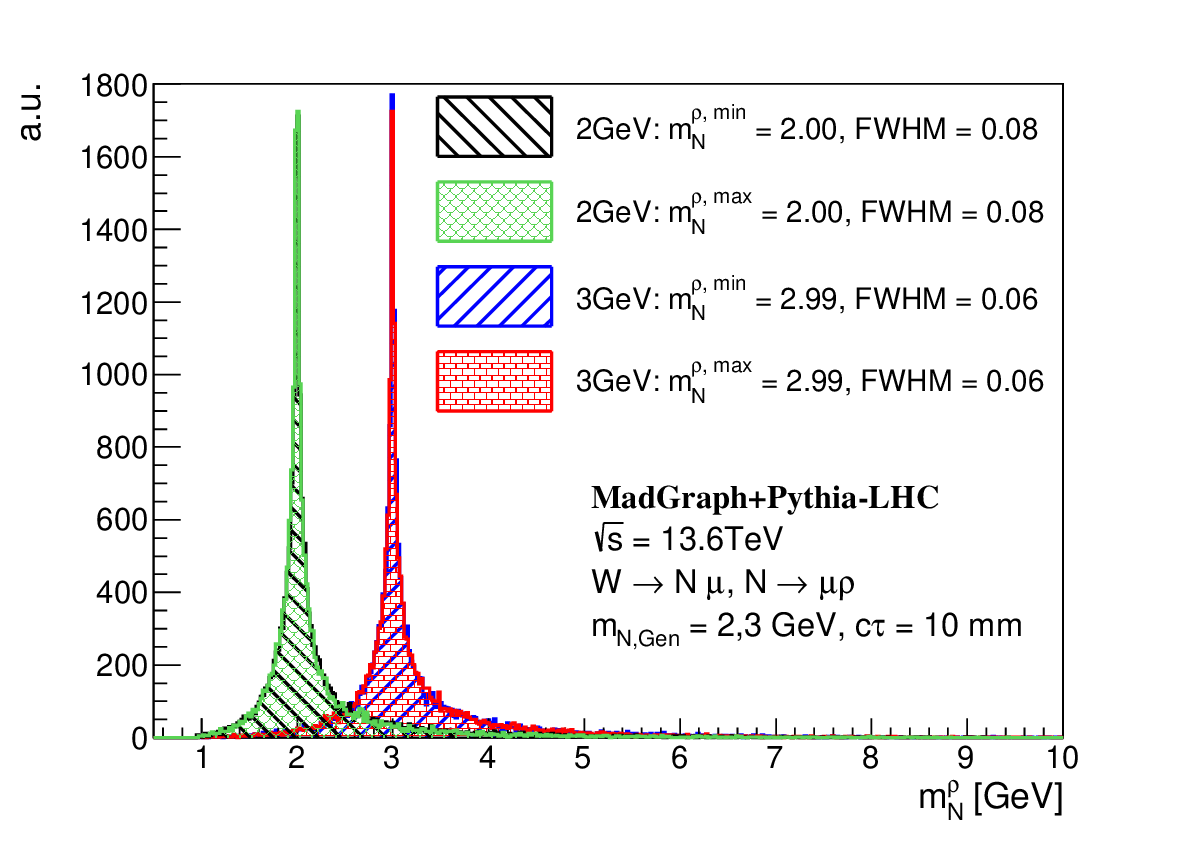}
    \caption{Comparison of reconstructed HNL mass using the \(\rho\)-mass constraint
method in an LHC-like sample (\(pp\!\to\!W^{\pm}\!+\!X\to\ell N\!+\!X\), \(N\!\to\!\ell\rho\)).
Overlaid distributions are shown for \(m_{N, Gen}=2~\mathrm{GeV}\) and
\(3~\mathrm{GeV}\). For each mass hypothesis, the two branches
\(m_{N}^{\rho,\mathrm{min}}\) and \(m_{N}^{\rho,\mathrm{max}}\) produce nearly identical spectra
peaking at the true value.}
    \label{fig:hnl_mass_rho_method_2vs3}
\end{figure}

The reconstructed $m_{N}^{\rho}$, computed for the two solutions, is 
shown in Figure~\ref{fig:hnl_mass_rho_method_2vs3} for generated HNL masses 
of $2$ and $3\,\mathrm{GeV}$. The MPV and corresponding FWHM of 
the $m_{N}^{\rho,\text{min/max}}$ distributions are summarized in 
Table~\ref{tab:rho_constraint_summary}. For each mass hypothesis, the two branches
\(m_{N}^{\rho,\mathrm{min/max}}\) produce nearly identical spectra
peaking at the true value.

\begin{figure}[htbp]
    \centering
    \includegraphics[width=0.53\textwidth]{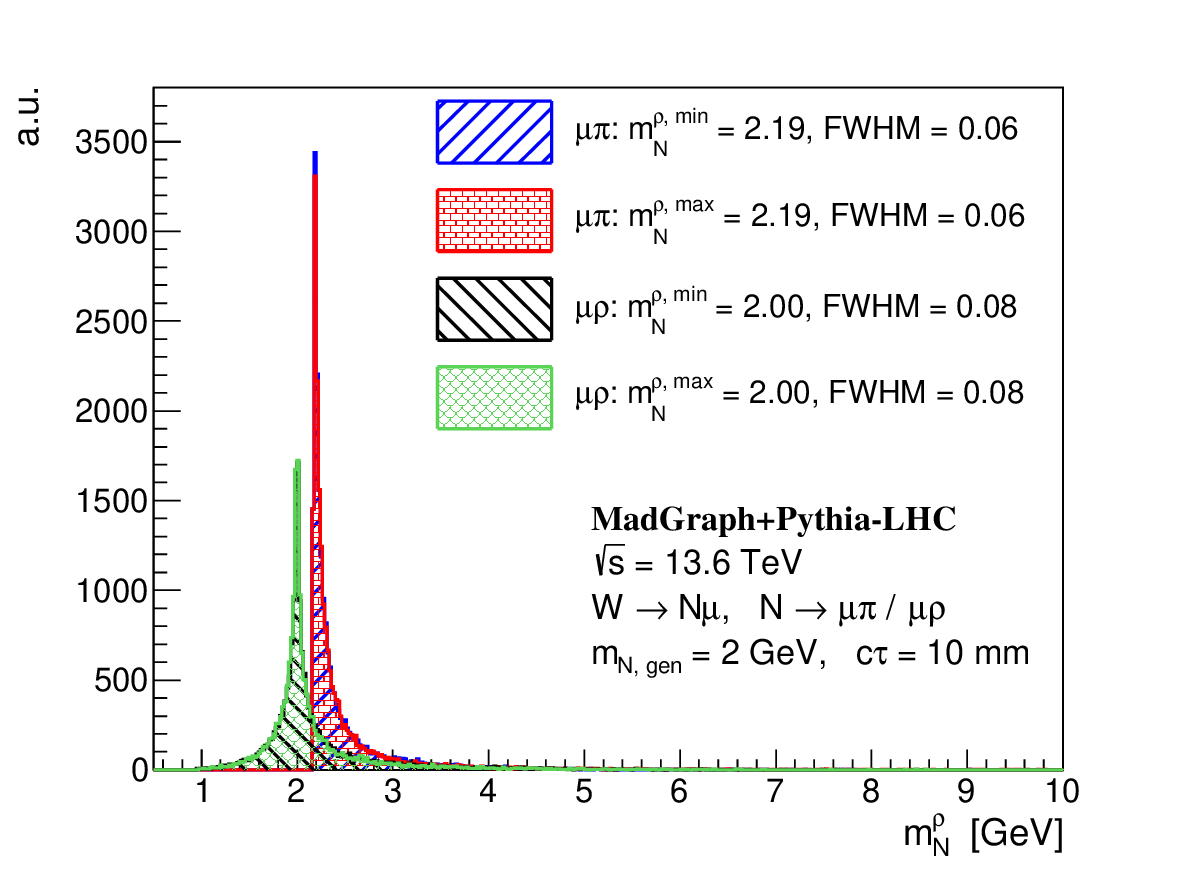}
    \caption{Comparison of reconstructed HNL mass using the \(\rho\)-mass constraint
method for the \(\mu\rho\) and \(\mu\pi\) channels in an LHC-like sample
(\(pp\!\to\!W^{\pm}\!+\!X\to\ell N\!+\!X\), \(N\!\to\!\ell\rho/\pi\)) with
\(m_{N,Gen}=2~\mathrm{GeV}\).
For \(N\!\to\!\mu\rho\), the two branches \(m_{N}^{\rho,\mathrm{min}}\) and
\(m_{N}^{\rho,\mathrm{max}}\) yield nearly identical spectra that peak at the true mass,
whereas for \(N\!\to\!\mu\pi\) the reconstructed \(m_{N}^{\rho}\) forms a displaced,
narrower peak, enabling clear separation of the decay hypotheses.}
    \label{fig:hnl_mass_rho_method_pivsrho}
\end{figure}

Figure~\ref{fig:hnl_mass_rho_method_pivsrho} compares the 
$m_{N}^{\rho}$ distributions for the $\mu\rho$ and $\mu\pi$ HNL deccay channels. 
The reconstructed \(m_{N}^{\rho}\) forms a displaced,
narrower peak for $\mu\pi$ HNL deccay, enabling clear separation of the decay hypotheses.
The observed peak separation reflects the method’s sensitivity to the 
underlying decay structure. At detector level, however, the achievable 
discrimination will be limited by the track--momentum resolution at the 
displaced vertex. For example, ATLAS reports percent-level dimuon mass 
resolution in the \(\mathrm{GeV}\) range for prompt, combined muons 
\cite{ATLASMuonCalib2023}; resolutions for displaced 
tracks are typically worse and analysis-dependent. Hence, while the 
particle-level peak offset is plausibly resolvable, a definitive 
assessment requires full detector simulation. 
n addition to the $m_{N}^{\rho}$ peak position and the impact parameters at the primary interaction vertex (similar to the SHiP case; see subsection 5.1), there are two other handles to distinguish the \(\ell\rho\) and \(\ell\pi\) HNL decay channels (in tracking-only reconstruction):

(i) \(m_{\ell_{\text{prompt}}+\text{DV}}\): the invariant mass of the
prompt lepton with the visible displaced system, which is 
systematically lower in the $\ell\rho$ channel due to the undetected 
$\pi^{0}$ carrying away energy.

(ii) \(m_{{\text{DV}}}\): the invariant mass of the displaced system. For \(N\!\to\!\ell\pi\) (with \({\text{DV}}=\ell+\pi^\pm\) fully
visible) this observable peaks at \(m_N\) up to resolution. For
\(N\!\to\!\ell\rho\) (with \({\text{DV}}=\ell+\pi^\pm\) and a missing
\(\pi^{0}\)) the distribution is broader and exhibits a kinematic endpoint at
\(m_N\) at particle level.

\end{sloppypar}
\begin{sloppypar}
\end{sloppypar}

\begin{table}[htbp]
\centering
\caption{The MPV of the reconstructed HNL mass and FWHM with the $\rho$-mass constraint method.}
\label{tab:rho_constraint_summary}
\scriptsize
\begin{tabular}{lllcc}
\hline
Scenario & $m_\text{N,Gen}$ & Channel (sol.) & MPV & FWHM \\
\hline
SHiP-like& $1\,\text{GeV}$ & $\mu\rho$ (min/max) & 1.00 & 0.15 \\
SHiP-like& $1\,\text{GeV}$ & $\mu\pi$ (min)  & 1.28 & 0.05 \\
SHiP-like& $1\,\text{GeV}$ & $\mu\pi$ (max)  & 1.28 & 0.06 \\
LHC-like& $2\,\text{GeV}$ & $\mu\rho$ (min/max)   & 2.00 & 0.08 \\
LHC-like& $3\,\text{GeV}$ & $\mu\rho$ (min/max)   & 2.99 & 0.06 \\
LHC-like& $2\,\text{GeV}$ & $\mu\pi$ (min/max)& 2.19 & 0.06 \\
\hline
\end{tabular}
\end{table}

\section{\(W\)-Mass Constraint Method}
\label{W-conts-method}
\begin{sloppypar}
In hadron collider experiments, a common production mechanism for the HNL is considerd 
via the decay of a weakly produced on-shell \(W\) boson, where the \(W\) boson 
decays into a prompt lepton and an HNL (\(W \to \ell N\)). Although the \(W\) boson 
is not reconstructed directly from its decay products in such scenarios, its 
well-known mass~\cite{pdg2022} \(m_W = 80.37 \mathrm{GeV}\) and width 
\(\Gamma_W = 2.19 \mathrm{GeV}\) can be utilized as a global kinematic 
constraint. By assuming that the prompt lepton and the HNL originate from a 
common \(W\) decay, the invariant mass of the system 
\(\ell_{\text{prompt}} + N\) can be constrained to match the known \(W\) boson mass. 
The Feynman diagram illustrating this process is shown in 
Figure~\ref{fig:W_to_N_decay}. 
\end{sloppypar}
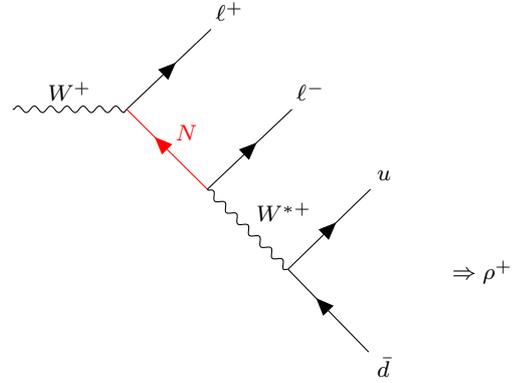
\begin{figure}[htbp]
\centering
\begin{tikzpicture}
  \begin{feynman}
    \vertex (w0); 
    \vertex [right=of w0] (w);
    \vertex [above right=of w] (lp) {\(\ell^+\)}; 
    \vertex [below right=of w] (hnl); 
    \vertex [above right=of hnl] (lm) {\(\ell^-\)}; 
    \vertex [below right=of hnl] (v2);
    \vertex [above right=of v2] (u) {\(u\)};
    \vertex [below right=of v2] (dbar) {\(\bar{d}\)};
    
    \diagram* {
      (w0) -- [boson, edge label=\(W^+\)] (w),
      (w) -- [fermion] (lp),
      (w) -- [anti fermion, red, edge label=\(N\)] (hnl),
      (hnl) -- [fermion] (lm),
      (hnl) -- [boson, edge label=\(W^{*+}\)] (v2),
      (v2) -- [fermion] (u),
      (v2) -- [anti fermion] (dbar),
    };
  \end{feynman}
  \node at ($(u)!0.5!(dbar) + (1.3,0)$) {\(\Rightarrow \rho^+\)};
\end{tikzpicture}
\caption{Production and decay of HNL: \(W^+ \rightarrow \ell^+ N\), followed by \(N \rightarrow \ell^- + W^{*+} \rightarrow \ell^- + u\bar{d}\). The \(u\bar{d}\) pair hadronizes into a \(\rho^+\).}
\label{fig:W_to_N_decay}
\end{figure}
\begin{sloppypar}
As it was mentioned in Section~\ref{sec:Kinematic} to further reconstruct the kinematics of the HNL decay, it is useful to exploit momentum conservation in an appropriately chosen reference frame, where we follow essentially the method used for \(pp\to\ell W, W\to\ell N, N\to\ell\ell'\nu\) described in~\cite{ATLAS2023}, by replacing the \(\nu\) by the \(\pi^0\) in the \(N\to\ell\rho, \rho\to\pi\pi^0\).
In the rotated frame the energy-momentum conservation relation can be written as:
\begin{equation}
    P_{W} = P_{\ell_{1}}+P_{\ell_{2}}+P_{\pi}+P_{\pi^0} = P_{\ell_1}+P_{\ell\pi}+P_{\pi^0}
    \tag{17}
    \label{eq:wmom_2}
\end{equation}
where the $P_{\ell_1}$ and $P_{\ell_2}$ are the four-momenta of prompt and displaced lepton respectively. For simplification the sum of the four-momenta of displaced lepton and the displaced pion is defined as $P_{\ell\pi} = P_{\ell_2}+P_{\pi}$.
Using Equation~\ref{eq:tranM} the \(E_{\pi^{0}}\) can be written as:
\begin{equation}
    E_{\pi^{0}} = \sqrt{m_{\pi^0}^2 + p_{\ell\pi}^{\perp 2} + q^{\parallel 2} }
    \tag{18}
    \label{eq:wmom}
\end{equation}
 Squaring both sides of Equation~\ref{eq:wmom_2} one obtains:
 \begin{equation}
     m_W^2 = m_{\ell_1}^2+m_{\pi^0}^2+m_{\ell\pi}^2+2P_{\ell_1}.P_{\ell\pi} + 2P_{\ell_1}.P_{\pi^0}+2P_{\ell\pi}.P_{\pi^0}
     \tag{19}
     \label{mw_2}
 \end{equation}
 The last two terms of Equation~\eqref{mw_2} can be written as:
\begin{equation}
\begin{aligned}
     2P_{\ell_1}\cdot(P_{\ell\pi}+P_{\pi^0}) &= 2E_{\ell_1}(E_{\ell\pi}+E_{\pi^0})-2p_{\ell_1}^{\parallel}(q^{\parallel }+p_{\ell\pi}^{\parallel}) \\
     \end{aligned}
     \tag{20}
 \end{equation}
\begin{equation}
\begin{aligned}
     2P_{\ell\pi}\cdot P_{\pi^0} = 2E_{\ell\pi}E_{\pi^0}- 2p_{\ell\pi}^{\parallel}q^{\parallel} +2p_{\ell\pi}^{\perp 2}
     \end{aligned}
     \tag{21}
 \end{equation}
Then Equation~\ref{mw_2} results in:
\begin{multline}
m_W^2 - m_{\ell_1}^2 - m_{\pi^0}^2 - m_{\ell\pi}^2 - 2E_{\ell_1}E_{\ell\pi} 
- 2p_{\ell\pi}^{\,\perp\,2} 
+ 2p_{\ell_1}^{\,\parallel}p_{\ell\pi}^{\,\parallel} \\
= 2E_{\pi^0}(E_{\ell_1} + E_{\ell\pi})
- 2\left( p_{\ell_1}^{\,\parallel} + p_{\ell\pi}^{\,\parallel} \right) q^{\parallel}
\tag{22}
\label{mw_2_2_1}
\end{multline}
The mass of the prompt lepton in the energy regime of interest, can be neglected, so that solving Equation~\ref{mw_2_2_1} for $E_{\pi^0}$ gives:
\begin{equation}
    E_{\pi^0} = A+B q^{\parallel}
    \tag{23}
    \label{eq:epi0}
\end{equation}
\begin{equation}
A =\frac{m_W^2 - m_{\pi^0}^2 - m_{\ell\pi}^2 - 2E_{\ell_1}E_{\ell\pi} 
- 2p_{\ell\pi}^{\,\perp\,2} 
+ 2p_{\ell_1}^{\,\parallel} p_{\ell\pi}^{\,\parallel} }{2(E_{\ell_1}+E_{\ell\pi})}
\tag{24}
\label{mw_2_2}
\end{equation}
\begin{equation}
    B = \frac{p_{\ell_1}^{\,\parallel}+p_{\ell\pi}^{\,\parallel} }{E_{\ell_1}+E_{\ell\pi}}\,.
    \tag{25}
\end{equation}
 Substituting Equation~\ref{eq:wmom} in Equation~\ref{eq:epi0} gives the following quadratic equation as a function in $q^{\parallel}$:
 \begin{multline}
q^{\parallel^2}(B^2-1) +2ABq^{\parallel} +A^2-m_{\pi^0}^2-p_{\ell\pi}^{\,\perp 2}=0
\tag{26}
\label{eq:alpha}
\end{multline}
with the solutions:
\begin{multline}
q^{\parallel}_{max/min} =\\
\frac{-AB \pm\sqrt{p_{\ell\pi}^{\,\perp 2}(B^2-1)+A^2+m_{\pi^0}^2(B^2+1)}}{B^2-1}\,.
\tag{27}
\label{eq:alpha}
\end{multline}

The $+$ and $-$ signs in front of the square root corresponds to 
\(q^{\parallel}_{\text{max}}\) and \(q^{\parallel}_{\text{min}}\) respectively.
In order to calculate the HNL mass, we square the four-momenta conservation relation:
\begin{equation}
    P_{N} = P_{\ell}+P_{\pi^0}+P_{\pi}
    \tag{28}
    \label{eq:Ncons}
\end{equation}
and obtain: 
\begin{equation}
    \begin{aligned}
        m_{N}^{2}&=m_{\ell}^2+m_{\pi}^2+m_{\pi^0}^2+2(E_{\pi}E_{\pi^0}-\vec{p}_{\pi}\cdot\vec{p}_{\pi^0}
        \\ &\quad+E_{\ell}E_{\pi}-\vec{p}_{\ell}\cdot\vec{p}_{\pi}+ E_{\ell}E_{\pi^0}-\vec{p}_{\ell}\cdot\vec{p}_{\pi^0})
    \end{aligned}
    \tag{29}
    \label{w-mN}
\end{equation}
Analogous to the \(\rho\)-mass constraint, the \(W\)-mass constraint yields two kinematic solutions for the HNL mass, denoted as \(m_{N}^{\text{W,min/max}}\) . The performance of both solutions is studied using MC-simulated events in Sec.~\ref{case-W}. In some events, however, no real solution exists because the radicand in Equation~\ref{eq:alpha} becomes negative. To address this, we use an adaptive \(W\)-mass treatment described in Section~\ref{sec:adaptiveMass}, in analogy to Ref.~\cite{ATLAS2023} for \(pp\to\ell W, W\to\ell N, N\to\ell\ell'\nu\).
\end{sloppypar}
\subsection{Adaptive Mass Technique for the \(W\)-Mass Constraint Method}
\label{sec:adaptiveMass}
\begin{sloppypar}
 The HNL is assumed to originate from the decay of an on-shell W boson, the mass of the W boson is expected to be consistent with its known value. However, due to its natural Breit-Wigner width ($\Gamma_{W}$), the reconstructed W mass varies event-by-event.
In the standard approach, a fixed nominal W mass is assumed, but this can lead to unphysical solutions when solving for the missing momentum, if the argument of the square root in the kinematic equations becomes negative. 
Instead, we treat \(m_W^2\equiv m_W^2(q^{\parallel})\) as a
function and minimize it with respect to \(q^{\parallel}\) (see
Equation~(\ref{mw_2_2_1})). This provides a lower bound on the possible W mass:
\begin{equation}
\begin{aligned}
    m_{W,\text{min}}^2 &= 
    m_{\pi^0}^2 + m_{\pi}^2 + \biggl(E_{\ell_1}(E_{\pi^0} +E_{\ell\pi})
    - p_{\ell_1}^{\parallel}p_{\ell\pi}^{\parallel} -\\
    &\quad E_{\ell\pi}E_{\pi^0} - 2(p_{\ell_1}^{\parallel}+p_{\ell\pi}^{\parallel})\frac{E_{\pi^0}(p_{\ell_1}^{\parallel}+p_{\ell\pi}^{\parallel})}{E_{\ell_1}+E_{\ell\pi}}\biggr).
\end{aligned}
\tag{30}
\label{W-mas-min}
\end{equation}
Initially, similar to the \(\rho\)-constraint method, the nominal \(W\) mass was replaced by the event-wise lower bound \(m_{W,\min}\) from Equation~\ref{W-mas-min}.
However, this did not eliminate all failures— in a non-negligible fraction of events the radicand of the quadratic in \(q^{\parallel}\) remained negative. To address this, an adaptive mass technique analogous to the strategy used in the ATLAS HNL leptonic-channel analysis with a \(W\)-mass constraint~\cite{ATLAS2023} is applied.
\end{sloppypar}
\begin{sloppypar}
The minimum mass from Equation~\ref{W-mas-min} is subsequently used as the lower integration bound in the probability density function (PDF) given by the relativistic Breit-Wigner form:
\begin{equation}
    P(m_{W}^{2}) \propto \frac{1}{(m_{W}^2 - m_{W,\text{nominal}}^2)^2 + (m_{W,\text{nominal}} \Gamma_{W})^2}.
    \tag{31}
    \label{eq:breit_wigner}
\end{equation}
Using Equation~\ref{eq:breit_wigner}, the cumulative distribution function (\(CDF\)) is expressed analytically as:
\begin{equation}
\begin{aligned}
    CDF(m_{W,\text{min}}^2) &= \frac{1}{\pi} \tan^{-1} \left( \frac{m_{W,\text{min}}^2 - m_{W,\text{nominal}}^2}{\Gamma_{W} m_{W,\text{nominal}}} \right) \\
    &\quad + \frac{1}{2}.
     \end{aligned}
     \tag{32}
\end{equation}

The $m_W^2(q^\parallel)$ grows without bound as
$|q^\parallel|\to\infty$ (\(E_{\pi^0}(q^\parallel)\ge m_{\pi^0}\)), so $m_{W,\max}^2=+\infty$ and therefore
$CDF(m_{W,\max}^2)=1$ by normalization of the cumulative distribution~\cite{PDG2024}.
To select a representative value of \( m_{W} \) consistent with this distribution and the kinematic bound,  a median probability value within the allowed range is defined as following:
\begin{equation}
\begin{aligned}
    p_{\text{med}} &= \frac{1 + CDF(m_{W,\text{min}}^2)}{2}, \\
    &\quad 
    \text{assuming } CDF(m_{W,\text{max}}^2) = 1.
    \end{aligned}
     \tag{33}
\end{equation}

The corresponding median \(W\) mass is then given by inverting the \(CDF\):
\begin{equation}
\begin{aligned}
    m_{W,\text{med}}^2 &= m_{W,\text{nominal}}^2 + \Gamma_{W} m_{W,\text{nominal}} \\
    &\quad \cdot \tan\left[\pi \left(p_{\text{med}} - \frac{1}{2}\right)\right].
    \end{aligned}
    \tag{34}
\end{equation}
By implementing this adaptive mass technique, we mitigate unphysical solutions that arise when fixing \(m_W\) to its nominal value. As an alternative, one may select the \(W\)-mass value event-by-event using a quantile-based (inverse–\(CDF\)) sampling of the relativistic Breit–Wigner line shape; see Cowan, Secs.~1.3 and 2.8, respectively~\cite{Cowan1998}.
 We do not adopt this option in our baseline, but it yields comparable results when the sampling range is physically truncated.

\end{sloppypar}
\subsection{Case Study: Validation of the \(W\)-mass Constraint Method}
\label{case-W}
\begin{figure}[htbp]
    \centering
    \includegraphics[width=0.53\textwidth]{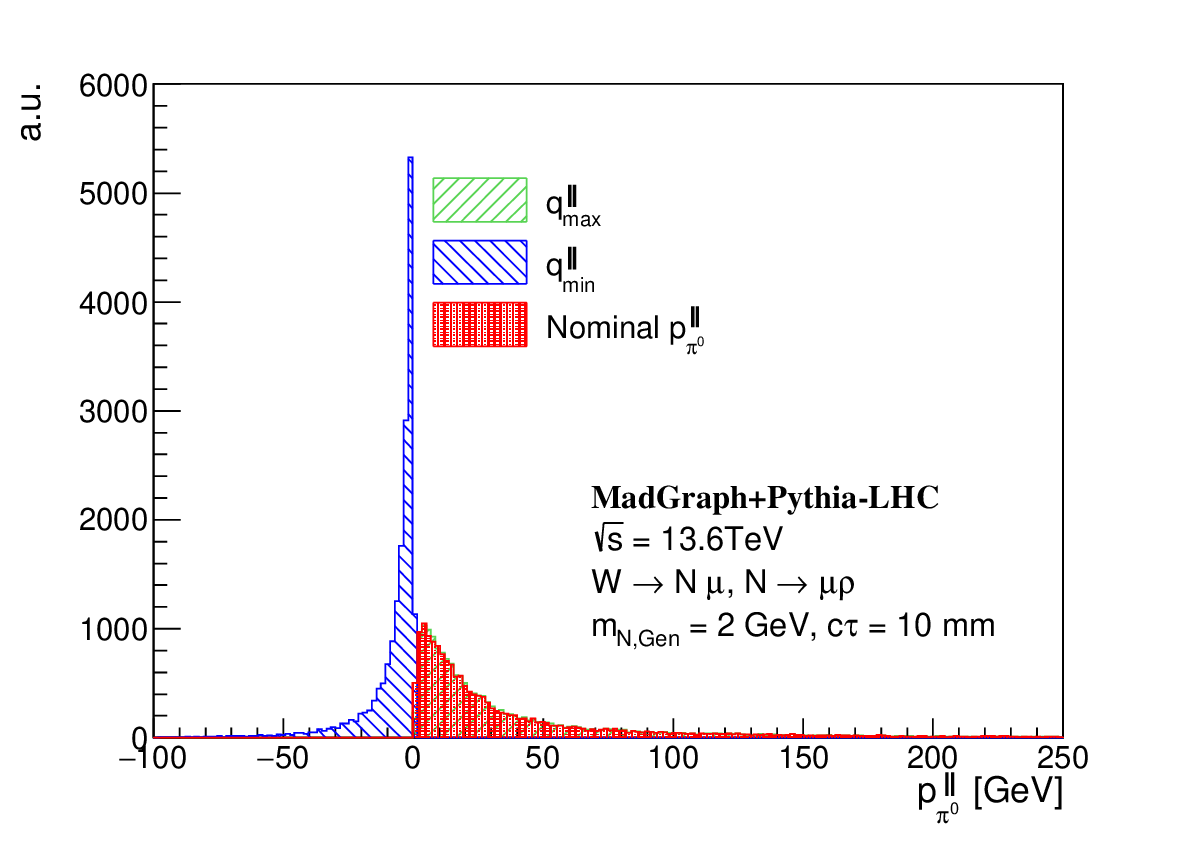}
    \caption{Longitudinal neutral–pion momentum \(p_{\pi^{0}}^{\parallel}\) for an LHC-like sample with \(m_{N,\mathrm{gen}}=2~\mathrm{GeV}\),
produced via \(pp\!\to\!W^{\pm}\!+\!X\to\ell N\!+\!X\) and \(N\!\to\!\ell\rho\).
The filled histogram shows the nominal (simulated value) \(p_{\pi^{0}}^{\parallel}\);
the hatched histograms show the two kinematic solutions \(q_{\text{min/max}}^{\parallel}\) obtained from the \(W\)-mass constraint method. The \(q_{\max}^{\parallel}\)
solution is closely follows the nominal distribution, whereas
\(q_{\min}^{\parallel}\) frequently yields unphysical values.}
    \label{fig:pp_pi0_1_W}
\end{figure}

\begin{figure}[htbp]
    \centering
    \includegraphics[width=0.53\textwidth]{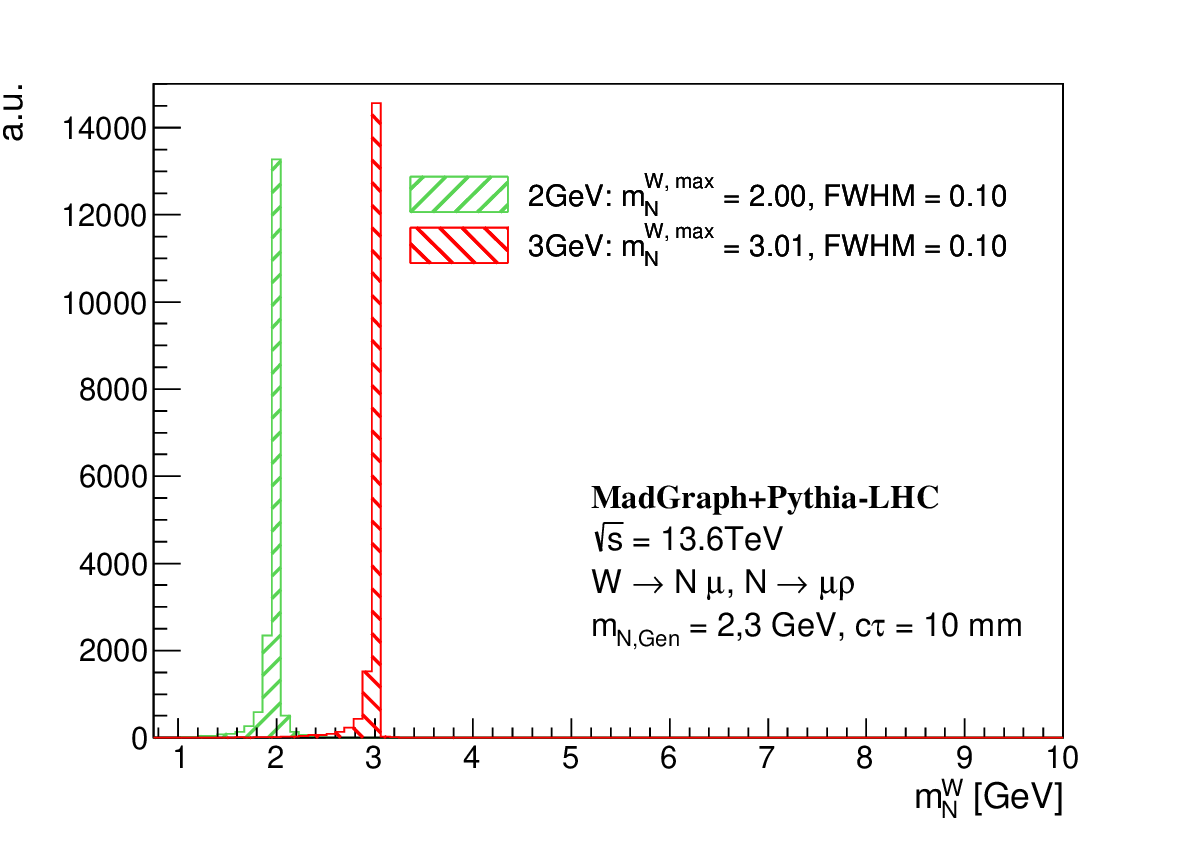}
    \caption{Comparison of reconstructed HNL mass \(m_{N}^{W,\text{max}}\) using the
\(W\)-mass constraint method in an LHC-like sample
(\(pp\!\to\!W^{\pm}\!+\!X\to\ell N\!+\!X\), \(N\!\to\!\ell\rho\)).
Overlaid spectra correspond to \(m_{N,Gen}=2~\mathrm{GeV}\) and \(3~\mathrm{GeV}\).}
    \label{fig:mHNL_overlay}
\end{figure}

\begin{sloppypar}
To validate the performance of the \(W\)-constraint reconstruction technique, a dedicated study was performed using a MC generated sample, as described in Section~\ref{sec:collidersim}.   
The distributions of \(p_{\pi^0}^{\parallel}\) obtained from the maximum and minimum solutions are shown in Figures~\ref{fig:pp_pi0_1_W}, together with the nominal (particle-level) value. The maximum solution yields non-negative values, whereas the minimum solution frequently becomes unphysical; therefore, only the maximum solution is used to reconstruct the HNL mass. Moreover, the maximum-solution distribution closely follows the nominal one, indicating that any reconstruction bias is small.
Figure~\ref{fig:mHNL_overlay} shows the reconstructed HNL mass, \(m_{N}^{W}\),
distribution obtained with the $W$-constraint method for generated HNL
masses of $2$ and $3\,\mathrm{GeV}$. The corresponding MPV values and 
FWHM are summarized in Table~\ref{tab:W_constraint_summary}. These 
results demonstrate the excellent mass resolution achievable with the 
$W$-constraint method, which significantly improves the separation of 
different HNL mass hypotheses.
\end{sloppypar}
\begin{figure}[htbp]
    \centering
    \includegraphics[width=0.53\textwidth]{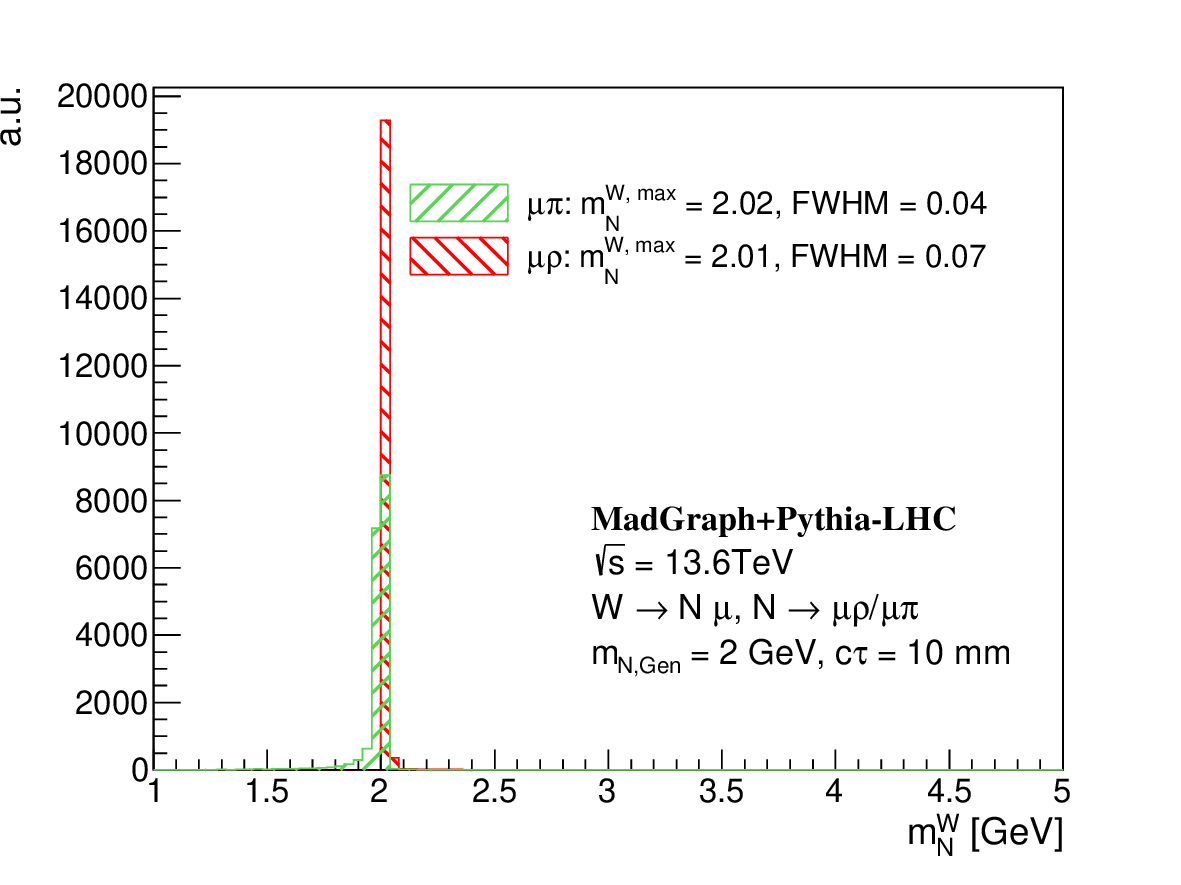}
    \caption{Comparison of reconstructed HNL mass \(m_{N}^{W,\text{max}}\) using the
\(W\)-mass constraint method for the \(\mu\rho\) and \(\mu\pi\) channels in an
LHC-like sample (\(pp\!\to\!W^{\pm}\!+\!X\to\ell N\!+\!X\), \(N\!\to\!\ell\rho/\pi\))
with \(m_{N,gen}=2~\mathrm{GeV}\).}
    \label{fig:mHNL_overlay_pion_rho}
\end{figure} 
\begin{sloppypar}
The \(N \rightarrow \mu \pi\) decay topology was generated and reconstructed using the same procedure. As shown in Figure~\ref{fig:mHNL_overlay_pion_rho}, unlike the \(\rho\)-mass constraint which enforces \(m(\pi^{+}\pi^{0}) \approx m_{\rho}\) and thus biases \(m_{N}\), the \(W\)-mass constraint relies solely on the production shell \((p_{l1}+p_{N})^2 = m_W^2\) and is therefore insensitive to the HNL decay products, yielding closer \(m_{N}\) values for both \(\mu+\rho\) and \(\mu+\pi\) event topologies.
\end{sloppypar}
\begin{table}[htbp]
\centering
\caption{The MPV of reconstructed HNL mass and FWHM with the $W$-mass constraint.}
\label{tab:W_constraint_summary}
\scriptsize
\begin{tabular}{lll c c}
\hline
Scenario & $m_N$&Channel (sol.) & MPV  & FWHM\\
\hline
LHC-like& $2\,\text{GeV}$ & $\mu\rho$ (max)  & 2.00 & 0.10 \\
LHC-like& $3\,\text{GeV}$ & $\mu\rho$ (max)  & 3.01 & 0.10 \\
LHC-like& $2\,\text{GeV}$ & $\mu\pi$ (max) & 2.02 & 0.04 \\
LHC-like& $2\,\text{GeV}$ & $\mu\rho$ (max)& 2.01 & 0.07 \\
\hline
\end{tabular}
\end{table}
\section{Outlook: Mass Reconstruction of HNLs Coupling to \(\tau\) Leptons}
\label{tau-coupling}
\begin{sloppypar}
Probing HNL couplings to $\tau$ leptons in HNL decays to final states with a tau lepton plus additional particles is compelling yet challenging. 
Unlike the electron or muon cases, the \(\tau\) must be reconstructed through its decays, which reduces sensitivity due to the \(\tau\) branching fractions to the used channels.
Moreover, for almost all subsequent $\tau$-lepton decays, leptonic or semileptonic, the reconstruction of a mass peak is not possible due to the missing momentum from the emitted neutrino(s).
A notable extension of the method applies for \(m_{N} > m_{\tau}+m_{\pi}\), namely
the channel \(N \to \tau^{+}\pi^{-}\) with \(\tau^{+}\to \pi^{+}\nu\). In this topology there is a
single invisible particle, so one can replace the \(\rho\)-mass constraint by a
\(\tau\)-mass shell constraint, 
\((P_{\nu}+P_{\pi^{+}})^{2} \;=\; P_{\tau}^{2} \;=\; m_{\tau}^{2}\) which is analogous in spirit but not identical in implementation to the
\(\rho\)-mass constraint used above. A variant with \(\tau^{+}\to K^{+}\nu\) is also possible,
but its branching fraction is substantially smaller due to CKM suppression
(\(\sim\!\lvert V_{us}/V_{ud}\rvert^{2}\)) and reduced phase space.

This \(\tau\) decay channel is expected to be very challenging for searches at the LHC, because the background of two-track DV candidates without a lepton in the DV is much higher than the ones with one or even two leptons (see e.g. ~\cite{ATLAS2019HNLW}). From this point of view, we expect the \(\tau\)-constraint mass reconstruction method of higher interest for a beam-dump experiment like SHiP, where the background level is expected to be much lower compared to the LHC conditions.

A complication comes from the fact that it is a-priori not known whether the process is \(N \to \tau^+ \pi^-\) or \(N \to \tau^- \pi^+\). As a result, one has to follow two branches in the analysis for these two hypotheses in order to check for a mass peak.

The reconstructed DV will be characterized by increased distance-of-closest-approach values between the two pion tracks
due to the finite $\tau$-lepton lifetime. With very precise tracking capabilities, this feature could help to distinguish on an event-basis which of the two opposite charged pions belongs more likely to the \(\tau\)-lepton decay. For a beam-dump experiment like SHiP, the momenta of the final state particles are typically at most of the order of several tens of \(\mathrm{GeV}\), implying an average \(\tau\) decay length of at most a few millimetres. This is likely too short to be resolved at a DV that is located downstream in the decay volume in a distance of up to 50 m away from the tracking detector. 

\end{sloppypar}

\section{Discussion and Conclusion}
\label{conclusion}
\begin{sloppypar}
In summary, the mass reconstruction methods for HNL decay \(N\rightarrow l\rho\) , \(\rho \rightarrow\pi\pi^0\) developed in this work rely on a rotated coordinate 
system aligned with the HNL flight direction, which allows the \(\pi^0\) 
momentum to be constrained by exploiting momentum conservation and known mass 
relations. This study also presents two different
HNL mass reconstruction based method. Both approaches were studied with particle-level simulated events, providing a clean test of 
the underlying kinematic reconstruction principles. It is demonstrated that the 
\(\rho\)-mass constraint method achieves good mass resolution in the SHiP 
experiment. This method, being independent of the production mechanism and 
relying solely on displaced-vertex final states, is particularly well-suited 
for low-background, fixed-target environments where HNLs emerge from meson 
decays. In collider experiments like the LHC, where HNLs are primarily 
produced via \(W\) boson decays, the \(W\)-mass 
constraint method add additional mass reconstruction capabilities for \(N\to\ell\rho\). 

At the LHC, combining the \(\rho\)-mass and \(W\)-mass constraints — when applicable — enhances search sensitivity by improving background rejection and enhance the mass reconstruction robustness, particularly in topologies where neutral pions missed the detection.
 For instance, in collider experiments with both prompt and displaced lepton signatures, utilizing both the \(W\) and \(\rho\)-mass constraint methods in a hybrid or sequential reconstruction strategy could offer improved discrimination against background processes that mimic semileptonic HNL decays.
It is important to note that our study did not include detector reconstruction effects. In a realistic experimental setup, finite momentum and vertex resolution are expected to degrade the reconstructed HNL mass resolution. This effect may be more pronounced for the W mass constraint method, which depends on the precise reconstruction of both the prompt-lepton and DV kinematics, compared to the \(\rho\) constraint method that only relies on the DV kinematics.
\end{sloppypar}

\begin{acknowledgements}
The authors would like to thank Martina Ferrillo and Maksym Ovchynnikov for their valuable support in generating the samples for the Fixed-Target studies using \texttt{EventCalc-SHiP}.
\end{acknowledgements}

\bibliographystyle{spphys}     
\bibliography{bibliography} 

\begin{thebibliography}{10}
\providecommand{\url}[1]{{#1}}
\providecommand{\urlprefix}{URL }
\expandafter\ifx\csname urlstyle\endcsname\relax
  \providecommand{\doi}[1]{DOI \discretionary{}{}{}#1}\else
  \providecommand{\doi}{DOI \discretionary{}{}{}\begingroup
  \urlstyle{rm}\Url}\fi

\bibitem{shaposhnikov2007}
M.~Shaposhnikov, Nuclear Physics B \textbf{763}(1–2), 49–59 (2007).
\newblock \doi{10.1016/j.nuclphysb.2006.11.003}.


\bibitem{esteban2020}
I.~Esteban, M.~Gonzalez-Garcia, M.~Maltoni, T.~Schwetz, A.~Zhou, Journal of
  High Energy Physics \textbf{2020}(9) (2020).
\newblock \doi{10.1007/jhep09(2020)178}.
\newblock \urlprefix\url{http://dx.doi.org/10.1007/JHEP09(2020)178}

\bibitem{gronau1984}
M.~Gronau, C.N. Leung, J.L. Rosner, Phys. Rev. D \textbf{29}, 2539 (1984).
\newblock \doi{10.1103/PhysRevD.29.2539}.


\bibitem{ATLAS2023}
ATLAS Collaboration, Physical Review Letters \textbf{131}(6) (2023).
\newblock \doi{10.1103/physrevlett.131.061803}.


\bibitem{CMS2022}
CMS Collaboration, Physical Review D \textbf{110}(1) (2024).
\newblock \doi{10.1103/physrevd.110.012004}.
\newblock \urlprefix\url{http://dx.doi.org/10.1103/PhysRevD.110.012004}

\bibitem{NA62HNL2021}
NA62 collaboration, Physics Letters B \textbf{778}, 137–145 (2018).
\newblock \doi{10.1016/j.physletb.2018.01.031}.
\bibitem{Alves_2025}
 G.F.S. Alves, P. S. Bhupal Dev, K.J. Kelly, P.A.N. Machado, Phys.
Rev. D 111, 015017 (2025). \newblock \doi{10.1103/PhysRevD.
111.015017}
\bibitem{Drewes2018}
M.~Drewes, J.~Hajer, J.~Klaric, G.~Lanfranchi, Journal of High Energy Physics
  \textbf{2018}(7) (2018).
\newblock \doi{10.1007/jhep07(2018)105}.
\newblock \urlprefix\url{http://dx.doi.org/10.1007/JHEP07(2018)105}

\bibitem{SHIP2016}
 S. Alekhin et al. Rep. Prog. Phys. 79(12), 124201 (2016). 
\newblock \doi{10.1088/0034-4885/79/12/124201}.

\bibitem{ATLAS2025}
ATLAS Collaboration, Journal of High Energy Physics \textbf{2025}(7) (2025).
\newblock \doi{10.1007/jhep07(2025)196}.
\newblock \urlprefix\url{http://dx.doi.org/10.1007/JHEP07(2025)196}

\bibitem{SHIP2019}
SHiP Collaboration.
\newblock Sensitivity of the ship experiment to heavy neutral leptons (2019).
\newblock \urlprefix\url{https://arxiv.org/abs/1811.00930}

\bibitem{Bondarenko2018}
K.~Bondarenko, A.~Boyarsky, D.~Gorbunov, O.~Ruchayskiy, Journal of High Energy
  Physics \textbf{2018}(11) (2018).
\newblock \doi{10.1007/jhep11(2018)032}.
\newblock \urlprefix\url{http://dx.doi.org/10.1007/JHEP11(2018)032}

\bibitem{pdg2022}
S.~Navas, et~al., Phys. Rev. D \textbf{110}(3), 030001 (2024).
\newblock \doi{10.1103/PhysRevD.110.030001}

\bibitem{EventCalcSHiP}
M.~Ovchynnikov,
\newblock SensCalc with EventCalc module: a Mathematica-based tool for simulating LLP decays in lifetime frontier experiments,
\newblock Zenodo, 2025.
\newblock \doi{10.5281/zenodo.15186139}.



\bibitem{pythia8}
T.~Sjöstrand, S.~Ask, J.R. Christiansen, R.~Corke, N.~Desai, P.~Ilten,
  S.~Mrenna, S.~Prestel, C.O. Rasmussen, P.Z. Skands, Computer Physics
  Communications \textbf{191}, 159–177 (2015).
\newblock \doi{10.1016/j.cpc.2015.01.024}.
\newblock \urlprefix\url{http://dx.doi.org/10.1016/j.cpc.2015.01.024}

\bibitem{SHiP:2022xqw}
C.~Ahdida, et~al., Eur. Phys. J. C \textbf{82}, 486 (2022).
\newblock \doi{10.1140/epjc/s10052-022-10346-5}

\bibitem{DeLellis:2018epjconf}
G.~De~Lellis, in \emph{{16th International Conference on Topics in
  Astroparticle and Underground Physics (TAUP 2019)}}, vol. 182 (2018), vol.
  182, p. 02016.
\newblock \doi{10.1051/epjconf/201818202016}

\bibitem{Ruiz}
R.~Ruiz, Physical Review D \textbf{103}(1) (2021).
\newblock \doi{10.1103/physrevd.103.015022}.

\bibitem{Madgraph}
J.~Alwall, R.~Frederix, S.~Frixione, V.~Hirschi, F.~Maltoni, O.~Mattelaer, H.S.
  Shao, T.~Stelzer, P.~Torrielli, M.~Zaro, Journal of High Energy Physics
  \textbf{2014}(7) (2014).
\newblock \doi{10.1007/jhep07(2014)079}.
\newblock \urlprefix\url{http://dx.doi.org/10.1007/JHEP07(2014)079}

\bibitem{PDG2024}
Particle Data Group,
\newblock Review of Particle Physics.
\newblock {\em Phys.\ Rev.\ D}, 110:030001, 2024.
\newblock \doi{10.1103/PhysRevD.110.030001}.
\newblock \url{https://pdg.lbl.gov/}.

\bibitem{NNPDF}
R.D. Ball, \textit{et al.}, JHEP \textbf{2015}(4), 040 (2015).
\newblock \doi{10.1007/JHEP04(2015)040}.


\bibitem{ATLASMuonCalib2023}
ATLAS Collaboration.
\newblock Eur.\ Phys.\ J.\ C \textbf{83}, 686 (2023).
\newblock \doi{10.1140/epjc/s10052-023-11584-x}.


\bibitem{Cowan1998}
G.~Cowan, \emph{Statistical data analysis} (Oxford University Press, USA, 1998)

\bibitem{ATLAS2019HNLW}
ATLAS Collaboration, Journal of High Energy Physics \textbf{2019}(10) (2019).
\newblock \doi{10.1007/jhep10(2019)265}.
\newblock \urlprefix\url{http://dx.doi.org/10.1007/JHEP10(2019)265}

\end{thebibliography}

\end{document}